\documentclass[showpacs,preprintnumbers,amsmath,amssymb]{revtex4}
\usepackage{graphicx}
\usepackage{bm}
\usepackage{amsfonts}
\usepackage{indentfirst}
\usepackage[small]{caption2}
\usepackage{longtable}

\begin{document}


\title{Squeezing and amplitude-squared squeezing effects on the dynamics of two nonidentical two-level atoms}
\author{E.K. Bashkirov}
 \altaffiliation[Electronic address:]{bash@ssu.samara.ru}
\affiliation{%
Department of General and Theoretical Physics, Samara State University, Acad. Pavlov Str.1 , 443011 Samara, Russia
\\
}%


 \begin{abstract}
Squeezing and amplitude-squared squeezing  for two two-level nonidentical atoms in lossless cavity has been investigated assuming the field to be initially in the coherent state. The time-dependent squeezing parameters has been calculated. The influence of the relative differences of two coupling constants on the squeezing parameters has been analyzed.
\end{abstract}



\maketitle

Squeezing phenomena attract much attention over the last few decades. The squeezed states of light were investigated
intensively both from theoretical and experimental point of view \cite{Scully} and attract considerable attention
because their possible practical applications for high-precision optical measurements,
 optical communications and optical processing \cite{Yamamoto}. A variety of  schemes for producing squeezed states
 has been proposed.

The possibility of squeezing phenomenon in Jaynes-Cummings model (JCM) was analyzed by several authors starting with
 the Meystre and Zubairy \cite{Meystre}. The multiphoton, nondegenerate two-mode  and two-atom generalizations of JCM
 have also been shown to produce squeezing \cite{Kien},\cite{Kien1}. The field squeezing in the two-atom JCM with one
 and multiphoton transitions has been investigated in several papers
 for initial coherent, squeezed, vacuum and thermal field input \cite{Kien1},\cite{Mir}. Last years some interest
 has arisen in higher-order squeezing \cite{Hong}. One type of higher-order squeezing, namely, squeezing of
 the square of the  field amplitude  or in brief the amplitude-squared squeezing (ASS)
  has been proposed by Hillery \cite{H}.
  The ASS has been shown to exist in one- and multiphoton JCM \cite{Yang} and two-atom JCM\cite{Mir},\cite{B}.

  In recent years, the model with two nonidentical two-level atoms in cavity has attracted a
  considerable attention in the study of the collective atom-field interaction.  The  exact solution of this model for lossless cavity and  the field which is at resonance with
  the atomic transitions
  has been calculated firstly for one-photon transitions by Zubairy et al. \cite{Z}, for two-photon transitions
   by Jex \cite{J} and for $m$-photon
 transitions by Xu et al. \cite{X}. Based on these solutions both the collapse-revival phenomenon of the
 atomic coherence for initial coherent \cite{Z},
 binomial \cite{S} and squeezed field state \cite{X} and the photon statistics
 \cite{X},\cite{A} have been considered. The entanglement of two nonidentical atoms, interacting with the
 thermal field
  in the cavity with loss has been studied in \cite{Zh}. Agarwal and co-authors have investigated the
  two-photon absorption \cite{K} and large two-photon vacuum
  Rabi oscillations \cite{P}
  in system of two nonidentical atoms taking into account the detuting. In this paper we consider the squeezing and ASS
in the system of the  two  atoms  with different coupling constants  which  interacts with one mode of coherent
 field in lossless cavity.
 We analyse the dependence of the squeezing on the relative difference of two coupling constants.

  Let us consider a system of two nonidentical two-level atoms interacting with a single-mode quantized electromagnetic
  field in a lossless resonant cavity via the one-photon-transition mechanism. The Hamiltonian of
  the considered system in the rotating wave approximation is
\begin{equation}
H = \hbar \omega a^+ a + \sum\limits_{i=1}^2 \hbar \omega_0 R^z_i + \sum\limits_{i=1}^2 \hbar g_i (R^+_i a + R^-_i a^+),\label{1}
\end{equation}
where $a^+$ and $a$ are the creation and annihilation operators of photons of the cavity field, respectively,
 $R^+_f$ and $R^-_f$ are the raising and the lowering operators for the $i$th atom, $\omega$ and $\omega_0$ are
 the frequencies of the field mode and the atoms, $g_i$ is the coupling constant between the $i$th atom and the field.
 We assume the field to be at one-photon resonance with the atomic transition, i.e. $\omega_0 = \omega$ .

We denote by  $\mid + \rangle$ and $\mid - \rangle$ the excited and ground states of single atom and by
 $\mid n\rangle$ the Fock state of the electromagnetic field. The two-atom wave function can be expressed as
 a combination of state vectors of the form $\mid \it v_1, \it v_2 \rangle = \mid \it v_1\rangle \mid \it v_2 \rangle $,
 where $\it v_1, \it v_2 = +,-$.
   Let  the atoms are initially in the ground state $\mid -, -\rangle$ and the field is initially in a coherent state
   $\mid \alpha \rangle$,
$$\mid \alpha \rangle = \sum\limits_{n=0}^{ \infty}\exp
 \left (-\frac{\mid \alpha \mid^2}{2}\right )
 \frac{\alpha^n}{\sqrt{n!}}, $$
 where $ \alpha = \mid \alpha \mid
 e^{\imath \varphi}$ and $ \overline{n} = \mid \alpha \mid^2$ is
  the initial mean photon number or dimensionless intensity of the  cavity field.

  The time-dependent wave function of the total system $\mid \Psi(t) \rangle$ obeys the ${\rm Schr\ddot{o}dinger}$ equation
$$
\imath \hbar  \mid \dot\Psi(t) \rangle = H \mid \Psi(t) \rangle. \eqno{(2)}
$$
Using the Hamiltonian (1) the wave function
  is found to be
 $$
 \mid \Psi(t) \rangle  =  \sum\limits_{n=0}^{ \infty} \exp[-\imath (n-1)\omega t]\exp
 \left (-\frac{\mid \alpha \mid^2}{2}\right )
 \frac{\alpha^n}{\sqrt{n!}}\times$$
$$ \times[ C^{(n)}_1(t) \mid + , +; n-2 \rangle + C^{(n)}_2(t) \mid + , -; n-1 \rangle + C^{(n)}_3(t) \mid - , +; n-1 \rangle +
   C^{(n)}_4(t) \mid - , -; n  \rangle ]. \eqno{(3)}$$

With the help of formulas (1)-(3) we can obtain  the equations of motion for probability  coefficients $C^n_i(t)$. These equations  must be written
 separately for $n=0, \>n=1$ and
 $n\geq 2$:
$$
\dot C^{(0)}_i = 0\quad(i=1,2,3,4); \eqno{(4)}
$$
$$
\dot C^{(1)}_1 = 0,\quad\dot C^{(1)}_2 = -\imath  g_1 C^{(1)}_4,\quad\dot C^{(1)}_3 = -\imath g_2 C^{(1)}_4,\quad\dot C^{(1)}_4 = -\imath (g_1 C^{(1)}_2+g_2 C^{(1)}_3); \eqno{(5)}
$$
and for $n\geq 2$
$$
\dot C^{(n)}_1 = - \imath (g_2 \sqrt{n-1} C^{(n)}_2 +  g_1 \sqrt{n-1} C^{(n)}_3,$$
$$\dot C^{(n)}_2 = - \imath (g_2 \sqrt{n-1} C^{(n)}_1 +  g_1 \sqrt{n} C^{(n)}_4,$$
$$\dot C^{(n)}_3 = - \imath (g_1 \sqrt{n-1} C^{(n)}_1 +  g_2 \sqrt{n} C^{(n)}_4,$$
$$ \dot C^{(n)}_4 = - \imath (g_1 \sqrt{n} C^{(n)}_2 +  g_2 \sqrt{n} C^{(n)}_3. \eqno{(6)}$$

For atoms initially prepared in their ground state we have the initial conditions for probability  coefficients
$$
C^{(n)}_4(0) =1,\quad C^{(n)}_1(0)=C^{n)}_2(0) = C^{(n)}_3(0) =0\quad (n=0,1,2,\ldots). \eqno{(7)}
$$
The solutions of Eqs. (4)-(6) with initial conditions (7) are found to be
$$
C^{(0)}_1(t) =  C^{(0)}_2(t)=C^{(0)}_3(t) = 0,\quad C^{(0)}_4(t) = 1; \eqno{(8)}
$$
$$
C^{(1)}_1(t) = 0,\quad C^{(1)}_2(t) = \frac {-\imath \sin(\sqrt{1+R^2}t)}{\sqrt{1+R^2}}, $$
$$ C^{(1)}_3(t) = \frac {-\imath R\sin(\sqrt{1+R^2}t)}{\sqrt{1+R^2}}, \quad C^{(1)}_4(t) = \cos(\sqrt{1+R^2}t) \eqno{(9)}
$$
and for $n\geq 2$
$$
 C^{(n)}_1(t) = \frac {2 R\sqrt{(n-1)n}}{\beta}[\cos(\lambda_+ t) - \cos(\lambda_- t)],$$
$$ C^{(n)}_2(t) = \frac {-4 \imath R^2 (n-1)\sqrt{n}}{\beta}
 \left \{\frac{\lambda_+^2 + (1-R^2) n}{\lambda_+[\beta-(1+R^2)]} \sin(\lambda_+ t) -
\frac{\lambda_-^2 + (1-R^2) n}{\lambda_-[\beta+(1+R^2)]} \sin(\lambda_- t)
 \right \},$$
$$ C^{(n)}_2(t) = \frac {-4 \imath R (n-1)\sqrt{n}}{\beta}
 \left \{\frac{\lambda_+^2 - (1-R^2) n}{\lambda_+[\beta-(1+R^2)]} \sin(\lambda_+ t) -
\frac{\lambda_-^2 - (1-R^2) n}{\lambda_-[\beta+(1+R^2)]} \sin(\lambda_- t)
 \right \},$$
$$ C^{(n)}_4(t) = \frac {8 R^2(n-1)n}{\beta}\left [\frac{\cos(\lambda_+ t)}{\beta-(1+R^2)} +
 \frac{\cos(\lambda_- t)}{\beta+(1+R^2)}\right ], \eqno{(10)}
$$
where
$$\lambda_{\pm} = \sqrt{(1+R^2)(2n-1) \pm \beta}/\sqrt{2},$$
$$ \beta =\sqrt{(2n-1)^2(1+R^2)^2 - 4 (n-1)n(1-R^2)^2}, \quad R=g_2/g_1.$$

To investigate the photon squeezing we introduce the two slowly varying quadrature components $X_1, X_2$ of field, defined
 by
 $$
  X_1=\frac{1}{2}(a e^{\imath \omega t} + a^+e^{-\imath \omega t}),$$
 $$ X_2=\frac{1}{2\imath}(a e^{\imath \omega t} - a^+e^{-\imath \omega t}).$$
  Thus
  $[X_1, X_2] = \imath/2$, which implies the uncertainty relation $(\Delta X_1)^2 (\Delta X_2)^2 \geq 1/16$, where
$ (\Delta X_i)^2  = \langle X_i^2 \rangle - \langle X_i \rangle^2 \quad (i=1,2)$ are variances of quadrature  components.
 Normal squeezing occurs when
 variances satisfy the relation $(\Delta X_i)^2 < 1/4 \quad (i= 1\, {\rm or}\, 2)$. The condition for
 squeezing one can write in the form
 $S_i < 0$, where squeezing parameters are
 \begin{eqnarray}
 \nonumber S_i = \frac{(\Delta X_i)^2) - 1/4}{1/4} = 4 (\Delta X_i)^2 - 1 \> \>(i=1,2).
 \end{eqnarray}
 The value $S_i=-1$ corresponds to 100\% squeezing in $i$th quadrature component.
 In terms of photon creation and annihilation operators we can rewrite squeezing parameters in the form
 $$
  S_1 = 2 \langle a^+ a \rangle + 2 Re \langle a^2 e^{2\imath \omega t} \rangle -
  4 (Re \langle a e^{\imath \omega t} \rangle)^2, \eqno{(11)}$$
 $$ S_2 = 2 \langle a^+ a \rangle - 2 Re \langle a^2 e^{2\imath \omega t} \rangle - 4 (Im \langle a e^{\imath \omega t} \rangle)^2.\eqno{(12)}
$$
Using (3)  we can obtain
$$
 \langle a^+ a \rangle = \overline{n} -  \left [2\sum\limits_{n=2}^{\infty} p_n \mid C^{(n)}_1 \mid^2 +
\sum\limits_{n=1}^{\infty} p_n (\mid C^{(n)}_2 \mid^2 + \mid C^{(n)}_3 \mid^2) \right ] = A_0,$$
$$ e^{\imath \omega t}\langle  a \rangle = \alpha \left \{\sum\limits_{n=2}^{\infty} p_n
(C^{(n)}_1)^*  C^{(n+1)}_1 \sqrt{\frac{n-1}{n+1}} + \sum\limits_{n=1}^{\infty} p_n [(C^{(n)}_2)^*  C^{(n+1)}_2 + \right. $$
$$\left.
 +(C^{(n)}_3)^*  C^{(n+1)}_3] \sqrt{\frac{n}{n+1}}+ \sum\limits_{n=0}^{\infty} p_n
 (C^{(n)}_4)^*  C^{(n+1)}_4  \right \}  = \alpha A_1,$$
$$ e^{2\imath \omega t}\langle  a^2 \rangle = \alpha^2 \left \{\sum\limits_{n=2}^{\infty} p_n
  (C^{(n)}_1)^*  C^{(n+2)}_1 \sqrt{\frac{(n-1)n}{(n+1)(n+2)}} + \sum\limits_{n=1}^{\infty} p_n[(C^{(n)}_2)^*  C^{(n+2)}_2 + \right. $$
$$ \left.
 \mbox{+} (C^{(n)}_3)^*  C^{(n+2)}_3] \sqrt{\frac{n}{n+2}} + \sum\limits_{n=0}^{\infty} p_n
 (C^{(n)}_4)^*  C^{(n+2)}_4  \right \} = \alpha^2 A_2 . \eqno{(13)}
 $$
 The parameter of initial coherent state is $\alpha = \sqrt{\overline{n}} \exp{i\varphi}.$ Let below $\varphi = 0$.
Then, for squeezing parameters $S_1$ and $S_2$ one can write
$$ S_1=2 A_0 + 2 \overline{n} A_2 - 4 \overline{n} A_1^2, \eqno{(14)}$$
$$ S_2=2 A_0 - 2 \overline{n} A_2 .\eqno{(15)}$$

 To define the squeezing of the square of the field amplitude  or amplitude-squared squeezing (ASS) we can introduce
 the quantities \cite{H}
 \begin{eqnarray}
 \nonumber Y_1=\frac{1}{2}(a^2 e^{2\imath \omega t} + a^{+2}e^{-2\imath \omega t}),\\
 \nonumber  Y_2=\frac{1}{2\imath}(a^2 e^{2\imath \omega t} - a^{+2}e^{-2\imath \omega t}).
\end{eqnarray}
The operators $Y_1$ and $Y_2$ correspond to the real and imaginary parts, respectively,
of the field amplitude squared and obey the commutation  relation $[Y_1,Y_2]= i(2 n+1)$,
where $n=a^+a$. The uncertainty relation for these two quantities has the form
 \begin{eqnarray}
 \nonumber (\Delta Y_1)^2 (\Delta Y_2)^2 \geq \langle n+1/2\rangle^2.
\end{eqnarray}
The ASS state in $Y_1$ exists if $(\Delta Y_2)^2 < \langle n+1/2\rangle$ and similarly for $Y_2$. Then, we
can introduce the squeezing parameters for ASS in the following form
$$
Q_i =\frac{(\Delta Y_i)^2 - \langle n+1/2\rangle}{\langle n+1/2\rangle} = \langle n+1/2\rangle^{-1}( (\Delta Y_i)^2 -1.
$$
The SSFA is obtained whenever $Q_i < 0$ for $i=1$ or $i=2$ and $Q_i = -1$ will correspond to 100\% SSFA. In terms of photon creation and annihilation operators we can rewrite SSFA squeezing parameters in the form \cite{B}
$$
  Q_1 = \frac{1}{4} \langle n+1/2\rangle^{-1} \left [2 \langle a^{+2} a^2 \rangle  +
  2 Re \langle a^4 e^{4\imath \omega t} \rangle -
  4 (Re \langle a^2 e^{2\imath \omega t} \rangle)^2\right ], \eqno{(16)}$$
 $$ Q_2 = \frac{1}{4} \langle n+1/2\rangle^{-1} \left [2 \langle a^{+2} a^2 \rangle  -
  2 Re \langle a^4 e^{4\imath \omega t} \rangle -
  4 (Im \langle a^2 e^{2\imath \omega t} \rangle)^2\right ].\eqno{(17)} $$
From  (3) we have

$$
 \langle a^{+2} a^2 \rangle =  \sum\limits_{n=4}^{\infty} p_n (n-2)(n-3) \mid C^{(n)}_1 \mid^2 +
+ \sum\limits_{n=3}^{\infty} p_n (n-1)(n-2)[ \mid C^{(n)}_2 \mid^2 +\mid C^{(n)}_3 \mid^2]+ $$
$$+\sum\limits_{n=2}^{\infty} p_n  \mid C^{(n)}_4 \mid^2  = A_3,$$
$$ e^{4\imath \omega t}\langle  a^2 \rangle = \alpha^2 \left \{\sum\limits_{n=2}^{\infty} p_n
  (C^{(n)}_1)^*  C^{(n+4)}_1 \sqrt{\frac{(n-1)n}{(n+3)(n+4)}} + \sum\limits_{n=1}^{\infty}
   p_n[(C^{(n)}_2)^*  C^{(n+4)}_2 + \right. $$
$$ \left.
 \mbox{+} (C^{(n)}_3)^*  C^{(n+4)}_3] \sqrt{\frac{n}{n+4}} + \sum\limits_{n=0}^{\infty} p_n
 (C^{(n)}_4)^*  C^{(n+24)}_4  \right \} = \alpha^24 A_4 . \eqno{(18)}
$$

With taking into account the  Eqs. (13),(16)-(18) we can rewrite the  ASS parameters $Q_1$ and $Q_2$ in the form
$$
  Q_1 = \frac{1}{4} \langle n+1/2\rangle^{-1} \left [2 A_3  + 2 \overline{n}^2 A_4 -4 \overline{n}^2 A_2^2\right ],\eqno{(19)}$$
 $$ Q_2 = \frac{1}{4} \langle n+1/2\rangle^{-1} \left [2 A_3  - 2 \overline{n}^2 A_4 \right ].\eqno{(20)} $$

Using the expressions (11)-(20) we have calculated the squeezing parameters $S_i$ and $Q_i$ for various initial
photon numbers $\overline{n}$ and relative differences of two coupling constant $R$.

Fig. 1 presents the long time behaviour of parameters $S_1$ and $S_2$ for $\overline{n}=0.2$ and $R = 0.5$.
For small field intensities $\overline{n}$
 as soon as $t >0$ we observe negative values of $S_1$ (squeezing in the first field quadrature component)
 and positive values of $S_2$. As times goes on, $S_1$ and $S_2$ start oscillating and reversing sign. The maximum
 degree of
 subsequent squeezing may be larger than that for the first squeezing. These features have much in common with
 that for the case of single or
 two identical atoms \cite{Meystre},\cite{Kien1}. With increasing of $\overline{n}$ the degree of squeezing
 in $S_1$ and the number of squeezing intervals
decreases.

Figs. 2-5 present the short time behaviour of squeezing parameter  $S_1$ (the first  squeezing) for different small
 field  input intensities $\overline{n}$ and  values of relative differences of two coupling constants.
 Obviously, that for case $R=0$ we have dealings with a single  atom and  the case $R=1 $ corresponds to two identical atoms.
 For small input intensity $\overline{n}$ (let's say  $0 \leq\overline{n}\leq 0.3 $) the degree of first squeezing increases with decreasing of $R$
 (as $R$ decreases from 1 to 0 the maximum obtainable degree of squeezing increases from 20\% to 27\%
 for $\overline{n}=0.2$). For field intensities
$\overline{n} \approx 0.3$ the maximum degree of squeezing is insensitive to choice of $R$.
 But for larger intensity input (lets say $\overline{n} >0.3$) the dependence of the degree of squeezing from $R$ is reversed.
When, for instance, $\overline{n} =0.4 $ the increasing of $R$  from 0 to 1 leads to increasing the  degree of squeezing from 18\% to 28\%.
Note that at the beginning of time scale the squeezing parameter $S_1$ for model with two nonidentical atoms takes the
 positive values in contrast to that for single atom or two identical atoms and the first squeezing of $S_1$
 is reached with some delay time. But this features is distinct only for relative large initial intensities. In Fig.4 we show
 the short time behaviour of squeezing parameters $S_1$ for models with $\overline{n} =0.8$ and different $R$.
 For $\overline{n} >0.8$ the $R$ -dependence of the  degree of squeezing  has nonmonotone character.  Note that
  for large input intensities, the parameter $S_1$ exhibits weak first squeezing and with increasing  $\overline{n}$
  the squeezing is vanished at first for intermediate values of $R$ (See Fig. 5).

 Fig. 6 presents the long time behaviour of ASS parameters $Q_1$ and $Q_2$ for $\overline{n}=0.2$ and $R = 0.5$.
 These parameters for small input intensity parameters are carried out in much the same way as $S_1$ and $S_2$
  but the amount of squeezing for
 ASS is less than that for second-order squeezing.  The maximum degree  of squeezing  in $Q_1$ decreases with increasing
 of the parameter $R$. The dependence $Q_1$ and $Q_2$ from  intensity $\overline{n}$ have the more complicated character
 but for large  intensities $\overline{n}$ the ASS is weak in both components.

  Figs. 7,8  present the short time behaviour of squeezing parameter  $Q_1$ (the first  ASS) for different
 field  intensities $\overline{n}$ and different values of relative differences of two coupling constants $R$.
For small input intensity $\overline{n}$ (let's say  $0 \leq\overline{n}\leq 0.7 $) the degree of first ASS
increases with decreasing of $R$
 (as $R$ decreases from 1 to 0 the maximum obtainable degree of squeezing increases from 5\% to 1.5\%
 for $\overline{n}=0.4$).  For $\overline{n} >0.7$ the $R$ -dependence of the  degree of squeezing  has nonmonotone
 character. In particular for model with $\overline{n} =0.8 $ the
  maximum of ASS is equal 6\% when $R=0.5$.
  Similarly to ordinary squeezing the
  first ASS  is appeared with some delay time when $0 < R <1$ and with increasing of the input
  intensity the ASS is vanished at first
  for intermediate values of $R$.

  Thus, we have considered the effects of squeezing and amplitude-squared squeezing of the cavity field mode in the
  model with two nonidentical atoms. The case in which the field is initially in a coherent state together with the
  atoms in the ground
  state has been examined. The long and short time behaviour of the squeezing and ASS parameters have
  been calculated.
The influence of the relative differences of two coupling constants on the squeezing parameters has
been analyzed. The investigation of the model with multiphoton transitions and other initial states of
field and atoms is the aim of our subsequent papers.

  This work was supported by RFBR grant
 04-02-16932.

\newpage
 \begin{figure}[!h]
\resizebox{90mm}{50mm}{\includegraphics{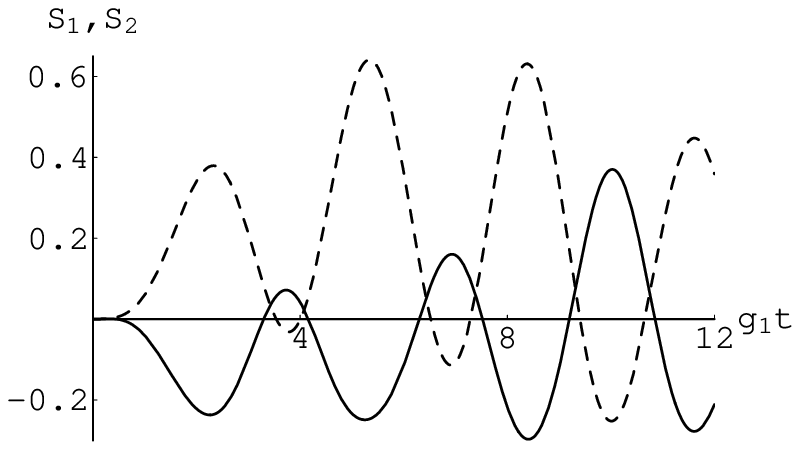}} \caption{Long time behaviour of the squeezing parameters
$S_1$ (solid line) and $S_2$ (dashed line)
 for model with $\overline{n}= 0.2$ and $R=0.5$.}
\end{figure}
\begin{figure}[!h]
\resizebox{90mm}{50mm}{\includegraphics{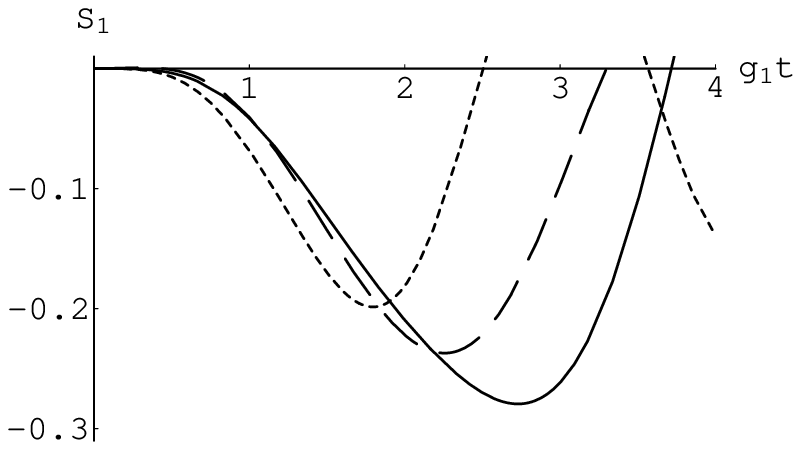}} \caption{Short time behaviour of the squeezing parameter $S_1$
 for model with $\overline{n}= 0.2$ and $R=0$ (solid line), 0.5 (dashed line)  and 1 (dotted line).}
\end{figure}
\begin{figure}[!h]
\resizebox{90mm}{50mm}{\includegraphics{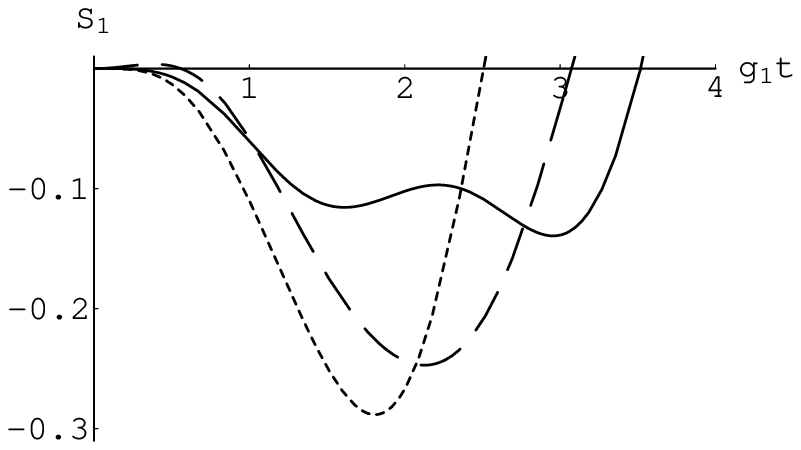}} \caption{Same as Fig. 2 but $\overline{n}= 0.4$.}
\end{figure}
\begin{figure}[!h]
\resizebox{90mm}{50mm}{\includegraphics{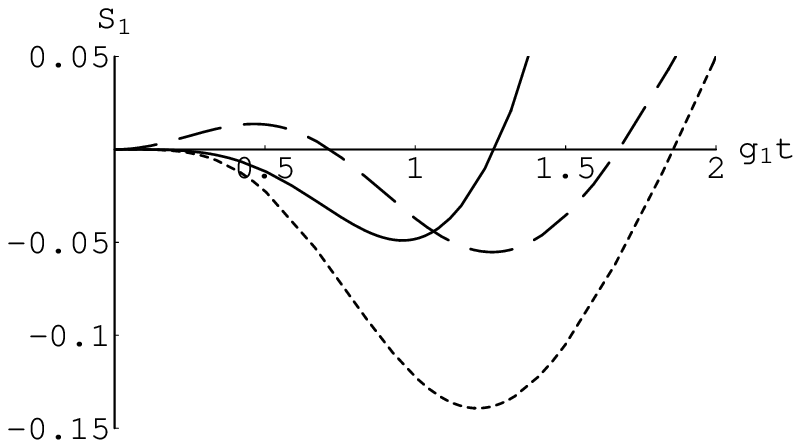}} \caption{Same as Fig. 2 but $\overline{n}= 0.8$.}
\end{figure}
\begin{figure}[!h]
\resizebox{90mm}{50mm}{\includegraphics{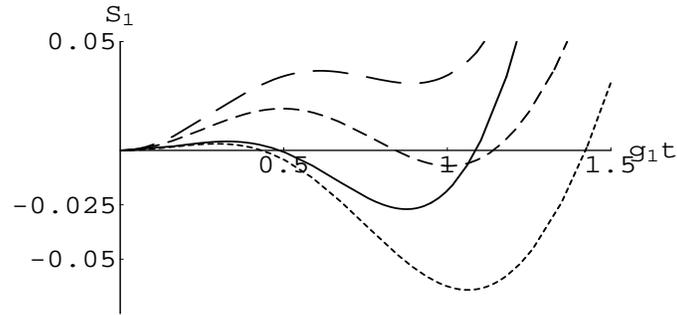}} \caption{Short time behaviour of the squeezing parameter $S_1$
 for model with $\overline{n}= 1.0$ and $R=0.1$ (solid line), 0.3 (dashed line), 0.5 (dashed line with small stroke)
   and 0.7 (dotted line).}
\end{figure}
\begin{figure}[!h]
\resizebox{90mm}{50mm}{\includegraphics{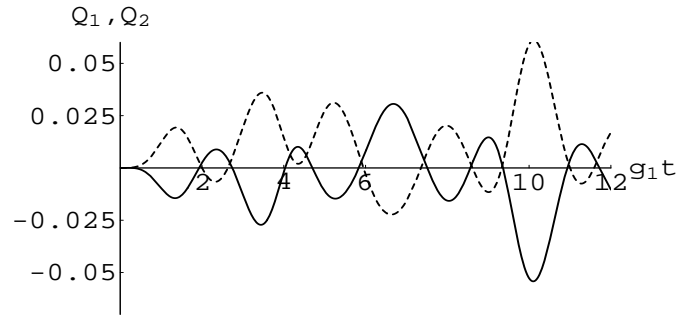}} \caption{Long time behaviour of the SSFA  parameters $Q_1$
(solid line) and $Q_2$ (dotted line)
 for model with $\overline{n}= 0.8
 $ and $R=0.5$.}
\end{figure}
\begin{figure}[!h]
\resizebox{90mm}{50mm}{\includegraphics{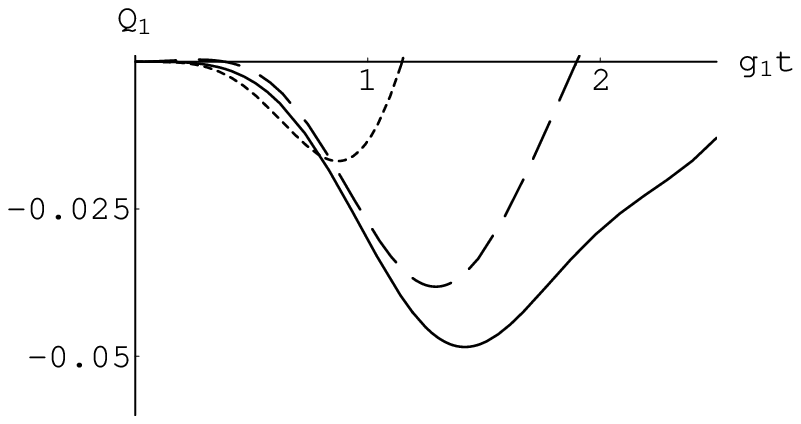}} \caption{Short time behaviour of the squeezing parameter $Q_1$
 for model with $\overline{n}= 0.4$ and $R=0$ (solid line), 0.5 (dashed line)  and 1 (dotted line).}
\end{figure}
\begin{figure}[!h]
\resizebox{90mm}{50mm}{\includegraphics{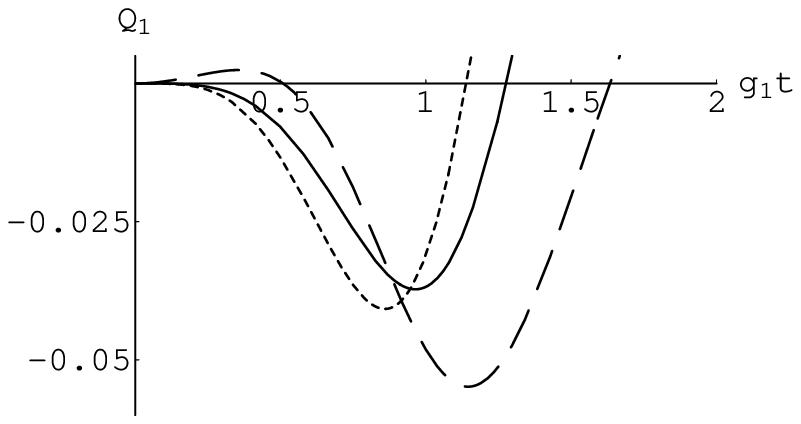}} \caption{Same as Fig. 7 but $\overline{n}= 0.8$.}
\end{figure}
\begin{figure}[!h]
\resizebox{90mm}{50mm}{\includegraphics{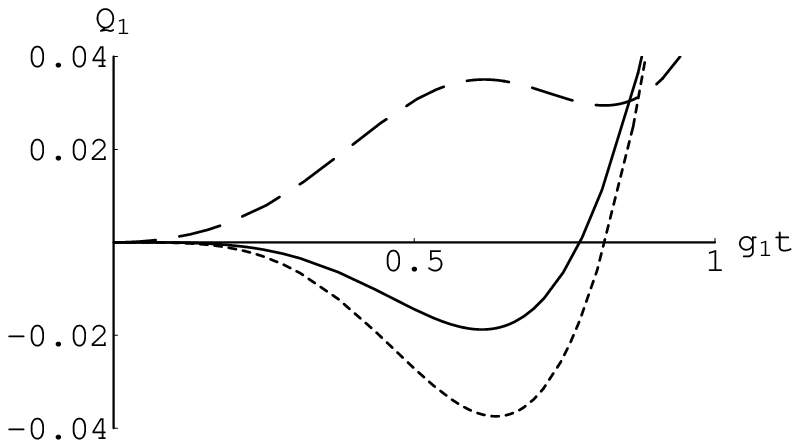}} \caption{Same as Fig. 7 but $\overline{n}= 1.2$.}
\end{figure}


\begin{thebibliography}{0}

\bibitem{Scully} M. O. Scully and M.S. Zubairy, {\it Quantum Optics}
(Cambrige University Press, 1997).

\bibitem{Yamamoto} Y. Yamamoto et al, in {\it Progress in Optics. Vol. 28}, ed. E. Wolf (Noth-Holland, Amsterdam, 1990), p.~89.

\bibitem{Meystre} P. Meystre and M.S. Zubairy, {\it Phys. Lett.} {\bf A89}, 390 (1982); M. Butler and P.D. Drummond,
{\it Optica Acta} {\bf 31}, 1 (1986);  P.L. Knight, {\it Physica Scripta T} {\bf 12}, 51 (1986);
G. Compagno, J.S. Pend and F. Persico, {\it Opt. Commun.} {\bf 57}, 415 (1986);
 K. Wodkiewicz et al, {\it Phys. Rev.} {\bf A35}, 2567 (1987); P.K. Aravid and G. Hu, {\it Physica} {\bf C150}, 427 (1988);
 S.M. Barnet and P.L.Knight, {\it Physica Scripta T} {\bf 21}, 5 (1988); J.R. Kuklinski and J.L. Madajczyk,
 {\it Phys. Rev.} {\bf A37}, 3175 (1988); M. Hillery, {\it Phys. Rev.} {\bf A39}, 1556 (1989).

\bibitem{Kien} A.S. Shumovsky, Fam Le Kien and E.I. Aliskenderov, {\it Phys. Lett.} {\bf A124}, 1987 (1987);
C.C. Gerry and P.J. Moyer, {\it Phys. Rev.} {\bf A38}, 5665 (1988); A. M. Abdel-Hafez, {\it Phys. Rev.} {\bf A45}, 6610 (1992).

\bibitem{Kien1} Z. Ficek, R. Tanas and S. Kielich, {\it Phys. Rev.} {\bf A29}, 2004 (1984); Fam Le Kien, E.P. Kadantseva and A.S. Shumovsky, {\it Physica}  {\bf C150}, 445 (1988);
Z.M. Zhang, L. Xu and J.-l. Chai, {\it Phys. Lett.}  {\bf A151}, 65 (1990).

\bibitem{Mir} M.A. Mir, {\it Intern. Journ. Mod. Phys.} {\bf B7}, 4439 (1993);
 M.A. Mir, {\it Intern. Journ. Mod. Phys.} {\bf B12}, 2743 (1998).

\bibitem{Hong} C.K. Hong and L. Mandel,  {\it Phys. Rev. Lett.} {\bf 54}, 323 (1985); C.K. Hong and L. Mandel,  {\it Phys. Rev.} {\bf A32}, 974 (1985).

\bibitem{H} M. Hillery,  {\it Opt. Commun.} {\bf 62}, 135 (1987).

\bibitem{Yang} X. Yang and X. Zheng, {\it Phys. Lett.} {\bf A138}, 409 (1989);
X. Yang and X. Zheng, {\it J. Phys.} {\bf B22}, 693 (1989); M. H. Mahran and A.-S.F. Obada, {\it Phys. Rev.} {\bf A40}, 4476 (1989);
 M. H. Mahran and A.-S.F. Obada, {\it Phys. Rev.} {\bf A42}, 1718 (1990);
 A. M. Abdel-Hafez and A.-S.F. Obada, {\it Phys. Rev.} {\bf A44}, 60717 (1991);
  M. A. Mir, {\it Phys. Rev.} {\bf A47}, 4384 (1993).

\bibitem{Mahrah} M.H. Mahrah and A.S.F. Obada, {\it Phys. Rev.} {\bf A40}; 4476 (1989); {\it Phys. Rev.} {\bf A42}, 1718 (1990).

\bibitem{B} E.K.Bashkirov and A.S. Shumovsky, {\it Intern. Journ. Mod. Phys.} {\bf B4}, 1579 (1990); E.K.Bashkirov et al,
 in {\it Photon echo and the problems of coherent optics --- Proc. IV Int. Symp. on Photon echo }, ed.~V. A. Katulin {\it et al.} (Samara State University, Samara, 1990), pp.~111--124.

\bibitem{Z} S. Mahmood and M.S.Zubairy, {\it Phys. Rev.} {\bf A35}, 425 (1987); S. Mahmood, K. Zaheer
 and M.S.Zubairy, {\it Phys. Rev.} {\bf A357}, 1634 (1988);  M.S.Iqbal, S.Mahmood, M.S.K.Razmi and M.S.Zubairy,
 {\it J. Opt. Soc. Am.} {\bf B2}, 1443 (1988).

\bibitem{J} I. Jex,  {\it Quantum Opt.} {\bf 2}, 443 (1990).

\bibitem{X} L. Xu, Z. Zhang and J.-L. Chai,  {\it J.Opt. Soc. Am.} {\bf 28}, 1157 (1991).

\bibitem{S} M.P. Sharma, D.A. Cardimona and A. Gavrielidies,  {\it J.Opt. Soc. Am.} {\bf B6}, 1942 (1989).

\bibitem{A} I. Ashraf and A.H.Toor,  {\it J.Opt.} {\bf B228}, 772 (2000).

\bibitem{Zh} X.X Yi, C.S. Yu, L. Zhou and H.S. Song,   {\it Phys. Rev.} {\bf A68}, 052304 (2003);
L. Zhou, X.X. Yi, H.S. Song and Y.Q. Quo,   {\it J. Opt.} {\bf B6}, 378 (2004).

\bibitem{K} M.S. Kim and G.S. Agarwal,   {\it Phys. Rev.} {\bf A57}, 3059 (1998).

\bibitem{P} P.K. Pathak and G.S. Agarwal,   {\it Phys. Rev.} {\bf A70}, 043807 (2004).

\end{thebibliography}
\end{document}